# Frame Selected Approach for Hiding Data within MPEG Video Using Bit Plane Complexity Segmentation

Hamid.A.Jalab, A.A Zaidan and B.B Zaidan

**Abstract**--- Bit Plane Complexity Segmentation (BPCS) digital picture steganography is a technique to hide data inside an image file. BPCS achieves high embedding rates with low distortion based on the theory that noise-like regions in an image's bit-planes can be replaced with noise-like secret data without significant loss in image quality. . In this framework we will propose a collaborate approach for select frame for Hiding Data within MPEG Video Using Bit Plane Complexity Segmentation. This approach will invent high secure data hidden using select frame form MPEG Video and furthermore we will assign the well-built of the approach; during this review the author will answer the question why they used select frame steganography. In additional to the security issues we will use the digital video as a cover to the data hidden. The reason behind opt the video cover in this approach is the huge amount of single frames image per sec which in turn overcome the problem of the data hiding quantity, as the experiment result shows the success of the hidden data within select frame, extract data from the frames sequence. These function without affecting the quality of the video.

───────── ◆ ─────────

## 1. INTRODUCTION

Steganography is the idea of hiding private or sensitive data or information within something that appears to be nothing out of the normal [1]. Steganography and cryptology are similar in the way that they both are used to protect important information. The difference between the two is that Steganography involves hiding information so it appears that no information is hidden at all [2]. If a person views the digital object that the information is hidden inside, he or she will have no idea that there is any hidden information, therefore the person will not attempt to decrypt the information, this is the main objective behind steganography [3]. Steganography comes from the Greek words Steganós (Covered) and Graptos (Writing), these days the sense of the word "Steganography" usually refers to information or a file that has been concealed inside a digital Picture, Video or Audio file [3],[4].What steganography technically does is to make use of human awareness; human senses are not trained to look for files that have information hidden inside of them[4].

───────────────────────

- Dr. Hamid.A.Jalab- Senior Lecturer, Department of Computer Science & Information Technology, University Malaya, Kuala Lumpur, Malaysia.
- A. A. Zaidan – PhD Candidate on the Department of Electrical & Computer Engineering, Faculty of Engineering, Multimedia University, Cyberjaya, Malaysia.
- B. B. Zaidan – PhD Candidate on the Department of Electrical & Computer Engineering / Faculty of Engineering, Multimedia University, Cyberjaya, Malaysia.

Although there are programs available that can do what is called Steganalysis (Detecting use of Steganography) [4].

The most common use of Steganography is to hide a file inside another file. When information or a file is hidden inside a carrier file, the data is usually encrypted with a password [4]. In this paper the researchers will focus on steganography on digital objects not Steganalysis, more specifically on digital video.

## 2. PREVIOUS WORKS

In this section I will review the main Steganography methods that have been used in video steganography and previous works been done based on those methods. Some if not most of these methods are also used by image steganography [5].

### 2.1 Steganography in Video Files Based on the YCbCr Color Space

YCbCr or Y'CbCr is a family of color spaces used as a part of the Color image pipeline in video and digital photography systems. YCbCr represents colors as a combination of three values:

- Y - The luminosity (roughly the brightness).
- Cb - the chrominance (roughly color) of the blue primary

- Cr - the chrominance (roughly color) of the red primary (Green is achieved by using a combination of these three values).

This technique is based on YCbCr. YCbCr space that can remove the correlation of R, G, and B in a given image, as less correlation between colors means less noticeable distortion. In this paper they concentrate on human video images, more specifically on human skin tones or colors for data hiding, note that human skin color can range from almost black to nearly colorless - appearing reddish white due to the blood vessels under the skin- in different people. The objective of this paper is to combat the use of forged passport documents or national identity cards, a security measure would be to embed individuals' information in their photos [6].

## 3. BIT PLANE COMPLEXITY SEGMENTATION

Bit Plane Complexity Segmentation steganography is a modern method of data hiding. Earlier methods of image steganography simply replaced the least significant bits of each pixel with hidden data [7]. This practice had very low embedding rates because visual defects rapidly develop as more significant bits are used. These defects are most noticeable in areas of homogenous color, where they usually appear as noise-like static. As more data is added, the noise [8].

Becomes more pronounced, until the image fades to unintelligible static. The degradation can become obvious and severe with only 10 to 15 percent of the image replaced with secret data. On computer and television screens, the smallest division of color data is a pixel. In computer memory, each pixel is represented by a binary value. The more bits that are used to represent each value, the wider the range of colors are for each pixel. Typical amounts of bits per pixel (bpp) are 8, 24, and 32. With these binary pixel values, and knowledge of which part of the picture each one represents, we can construct bit planes [9].

A bit plane is a data structure made from all the bits of a certain significant position from each binary digit, with the special location preserved. In Fig 1. position (0, 0) from bit plane 2 is bit 2 from pixel (0, 0) in the image. BPCS addresses the embedding limit by working to disguise the visual artifacts that are produced by the steganographic process [10].

Optometric studies have shown that the human visual system is very good at spotting anomalies in areas of homogenous color, But less adept at seeing them in visually complex areas. When an image is deconstructed into bit planes, the complexity of each region can be measured. Areas of low complexity such as homogenous color or simple shapes appear as uniform areas with very few changes between one and zero [11].

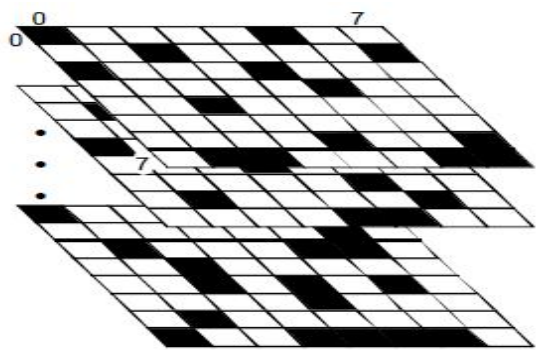

Fig. 1. Image pixel location (0, 0) has the binary value 01001110. In these bit planes, black is a 0 and white is a 1. In the first bit plane in the figure, position (0, 0), there is a black zero. In the second bit plane, there is a white one, and so on down to the last bit plane.

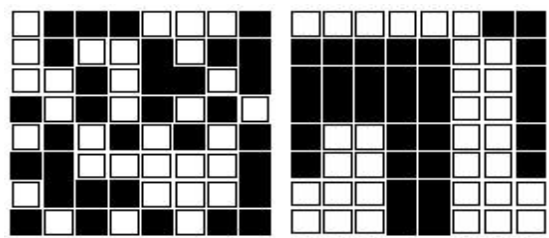

Fig. 2. Noise-like patch (a) and informative patch (b): (a) complexity 69, (b) complexity 29.

Complex areas such as a picture of a forest would appear as noise-like regions with many changes between one and zero. These random-seeming regions in each bit plane can then be replaced with hidden data, which is ideally also noise-like. Because it is difficult for the human eye to distinguish differences between the two noise-like areas, we are able to disguise the changes to the image. Additionally, since complex areas of an image tend to be complex through many of their bit planes, much more data can be embedded with this technique than with those that are limited to only the lowest planes.       In BPCS, the complexity of each subsection of a bit plane is defined as the number of non edge transitions from 1 to 0 and 0 to 1, both horizontally and vertically[12].

Thus the complexity of each section is not determined only by the number of one's or zeros it contains. Generally, for any square of 2nx2n pixels, the maximum complexity is 2x2nx(2n-1) and the minimum is of course 0. Most versions of image BPCS use an 8 pixel square, where the maximum complexity is 112. In Figure 2, white represents a one and black a zero. Both squares, and 'patches', have the same number of ones and zeros, but very different complexities. This shows that one contains much more visual information than the other. The complex patch (A) has very little visually informative information; therefore it can be replaced with secret date and with a very low effect on the image's quality. However, if the more visually informative patch (B) was replaced, it would cause noise-like distortion of the definite edges and shapes. This technique works very well with natural images, as they tend to have many areas of high complexity. Images with many complex textures and well shaded objects are usually having a high embedded data capacity [13]. BPCS works much less well with computer generated images and line art, as those classes of images tend to have large areas of uniformity and sharply defined border areas. With these types of images, there is very little complexity to exploit and any changes tend to generate very obvious artifacts [14].

This is one flaw BPCS shares with traditional steganography, though for slightly different reasons. Traditional steganography works poorly with computer generated pictures because the static distortion effect produced by embedding is very obvious in areas of homogenous color [15].

## 4. SYSTEM OVERVIEW

The main goal of our plan is to build a system program that is able to hide data in digital video files, more specifically in the images or frames extracted from the digital video file MPEG; as shown in Fig. 3.

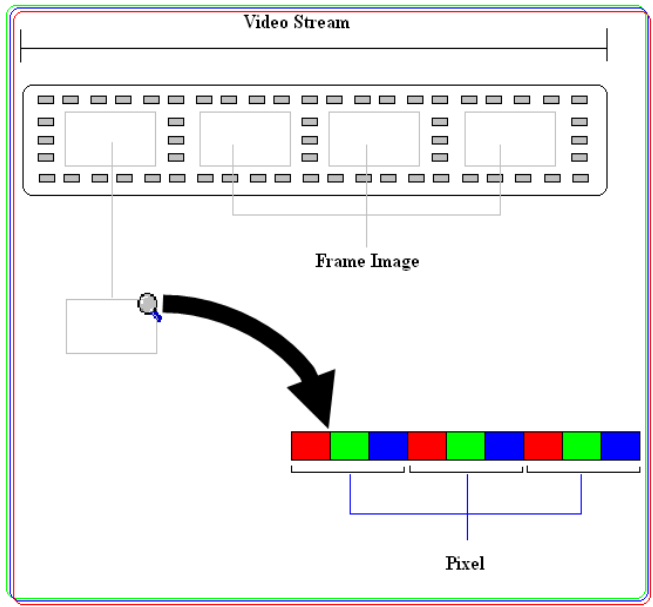

Fig. 3. Extracting Frames From Video File

The main function in this framework is steganography this approach carry out the dreamily protection for the information and make the attackers dream on getting data back. The algorithm work as the chart shows below, where unsuspected carrier with the strongest for select frame building the characteristic of our framework. The main function of the proposed approach is:

- Read Frames.
- Select Frame.
- Hidden the Data.
- Read Frames Sequence.
- Extract the Data.

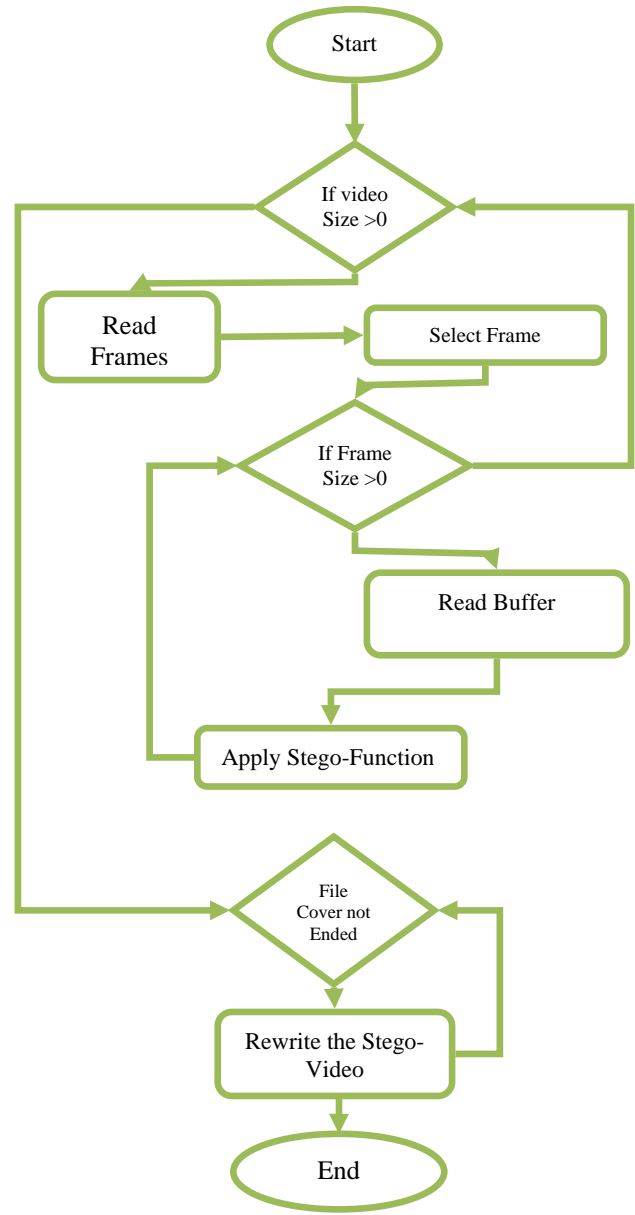

Fig. 4. The Encode Algorithm

The figure above showing the select frame with hidden operation this framework give more flexibility to appoint the start point at which frame as well the end point, this new feature make the system more secure in term of avoiding discover the data hidden using the statistical techniques. Fig.5. is the extraction operation with the select frame

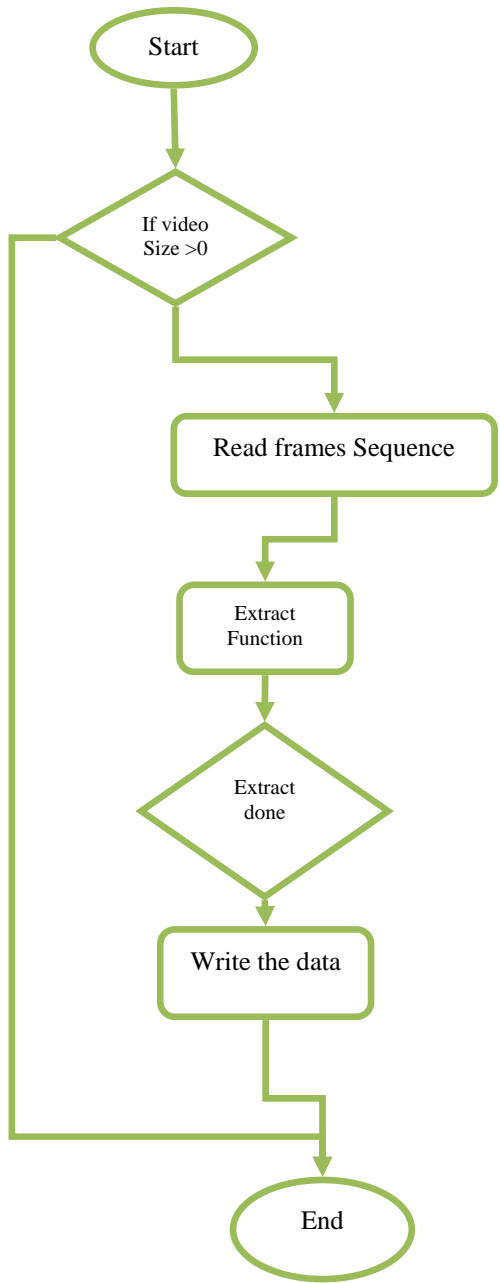

Fig. 5. Decoding Algorithm

## 5. EXPIREMENTAL RESULTS

Due to the difficulty of showing the result as a video stream on paper, the author prefers to display the result on the frame of the digital video file along with histogram of each a single frame. The following here are extracted frames of a digital video file. Fig. 6. Shows the frames from the famous movie "The Godfather" before applying the algorithm, while Fig. 7. Shows the frame after applying the algorithm. We can see here that there are no much differences between the two sets of frames especially for human vision system. This can tell that the algorithm can be applied successfully on video frames also to verify the algorithm by the histogram, to see the divergences on the frames before and after hiding data. From the histogram for both single frames in Fig. 6 & 7, its clear there is no differences between the two sets before and after hiding data which prove that the algorithm successfully hid the data into the frames without making a noticeable difference for the human vision system.

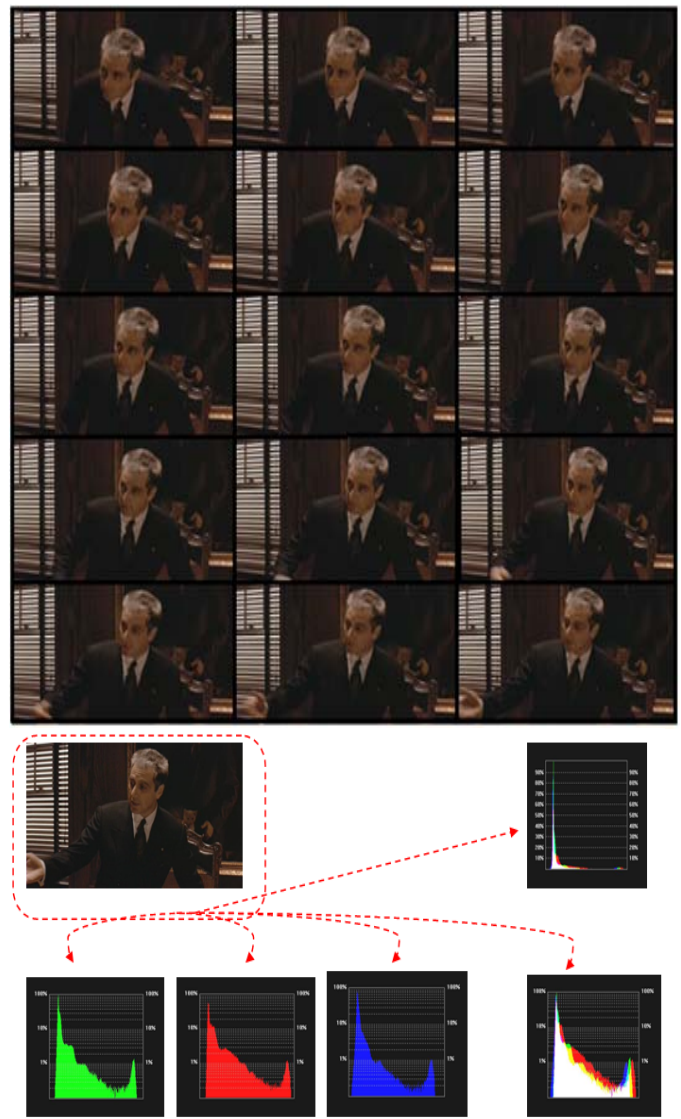

Fig. 6. Fifteen image frames has taken from a much known movie. "Godfather" before any hidden operation, the first frame under the histogram also the three channels on RGB has been separated for more accuracy on the test.

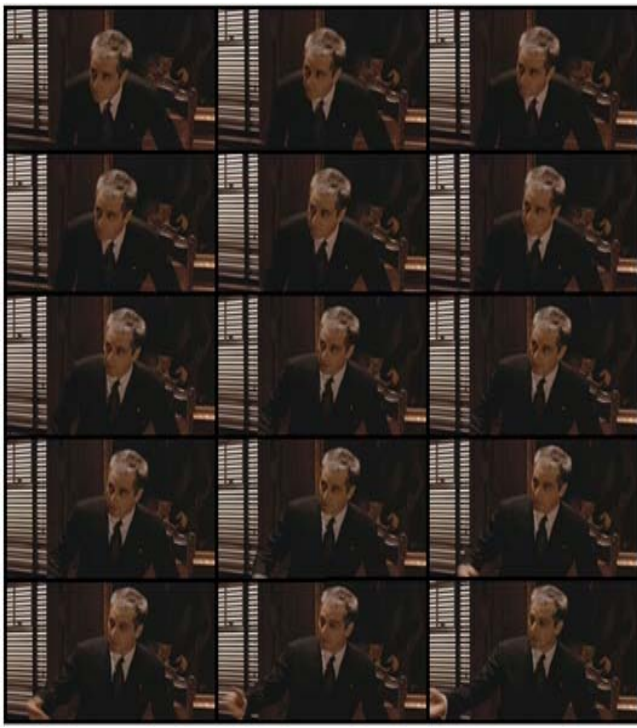

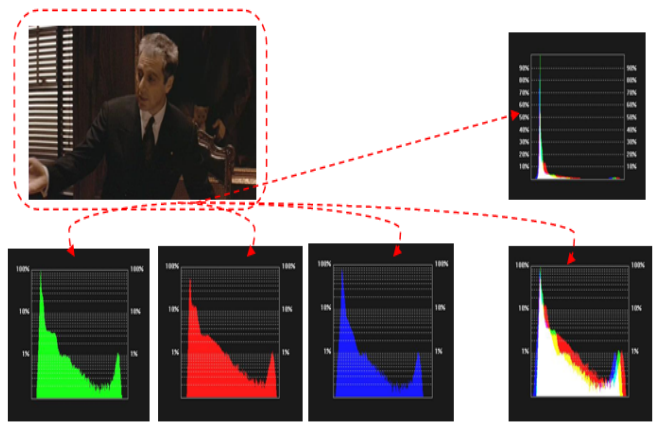

Fig. 7. Fifteen image frames has taken from a very known movie "Godfather" after hidden operation, the first frame under the histogram also the three channels on RGB has been separated,

## 6. CONCLUSION

In this paper, a new Approach of high secure video steganography has been invented. The basis of this method is use the digital video as separate frames and select frame to hides the information inside. As the experiment result shows the success of the hidden data within select frame, extract data from the frames sequence, these functions without affecting the quality of the video. This framework overcome the defeat of the limitation of steganography approach by invited the biggest size cover file among the multimedia file which is the video. In the video steganography we have a flexibility of make a selective frame steganography to higher the security of the system or using the whole video too high a huge amount of data hidden. Due the security issues the author has select frame from the whole frames which is in buffer, this idea make to guarantee the protection of data.

## ACKNOWLEDGEMENT

This work was supported in part by the University of Malaya, Kuala Lumpur Malaysia.

## REFERENCES

[1] A.A.Zaidan, B.B.Zaidan, "Novel Approach for High Secure Data Hidden in MPEG Video Using Public Key Infrastructure", International Journal of Computer and Network Security, Vol.1, No.1, P.P 71-76, 30 October, Vienna, Austria, 2009. (Journal paper).

[2] Alaa Taqa, A.A Zaidan, B.B Zaidan ,"New Framework for High Secure Data Hidden in the MPEG Using AES Encryption Algorithm", International Journal of Computer and Electrical Engineering (IJCEE),Vol.1 ,No.5, ISSN: 1793-8198, pp.589-595 , Singapore.,December (2009) . (Journal paper).

[3] Mohamed Elsadig Eltahir, Laiha Mat Kiah, B.B.Zaidan and A.A.Zaidan," High Rate Video Streaming Steganography", International Conference on Information Management and Engineering (ICIME09), Session 10, P.P 550-553, 2009 . (Conference proceeding).

[4] Fazida.Othman, Miss.Laiha.Maktom, A.Y.Taqa, B.B.Zaidan, A.A.Zaidan, "An Extensive Empirical Study for the Impact of Increasing Data Hidden on the Images Texture", International Conference on Future Computer and Communication (ICFCC 09), Session 7, P.P 477-481, 2009.(Conference proceeding).

[5] Hamid.A.Jalab, A.A Zaidan, B.B Zaidan, "New Design for Information Hiding with in Steganography Using Distortion Techniques", International Journal of Engineering and Technology (IJET)), Vol 2, No. 1, ISSN: 1793-8236, Singapore , Feb (2010), (Pending publication).

[6] A.A.Zaidan, B.B.Zaidan, Hamid.A.Jalab," A New System for Hiding Data within (Unused Area Two + Image Page) of Portable Executable File using Statistical Technique and Advance Encryption Standard ", International Journal of Computer Theory and Engineering (IJCTE), 2010, VOL 2, NO 2, ISSN:1793-8201,Singapore,2009, (Journal paper).

[7] Stallings, William, Network Security Essentials, applications and Standards p.75, Pearson Education, Inch, 2007.( Technical report with report number)

[8] Yeob , Chan & Farnham , Tim , (2003). Secure M-Commerce with WPKI, Toshiba Research Europe Limited, Toshiba Telecommunication Research Laboratory, available at: http://www.iris.re.kr/iwap01/program/download/g07_paper.pdf, (accessed May 4, 2008).

[9] Kuhn , D. Richard & Hu ,Vincent C. & Polk , W. Timothy & Chang , Shu-Jen , (2001). Introduction to Public Key Technology and the Federal PKI Infrastructure, National institute of standars and technology,(NIST),available,from:http://csrc.nist.gov/publications/nistpubs/800-32/sp800-32.pdf. (Accessed April 28, 2008).

[10] E. Kawaguchi, M. Niimi. "Modeling Digital image into informative and Noise-Like Regions by a Complexity Measure", 7th European-Japanese Conference on Information Modeling and knowledge Bases, 1997.( Conference proceeding)

[11] M. Niimi, H. Noda, E. Kawaguchi. "A Steganography Based on Region Segmentation by Using Complexity Measure". Trans. Of,IEICE, Vol. J81-D-II, No. 6, pp. 1132-1140, 1998.( Transaction, journal)

[12] R. Ouellette, H. Noda, M. Niimi, E. Kawaguchi. "Topological Ordered Color Table for BPCS Steganograpy Using Indexed Color Images", IPSJ Journal, 2000, (Journal paper).

[13] E. Kawaguchi, R.O. Eason. "Principal and Applications of BPCS Steganography", Proc of SPIE, Vol. 3528, pp. 464-473, 1998.. (Conference proceeding)

[14] J. M. Shapiro. "Embedded Image Coding Using Zerotrees of Wavelet Coefficients", IEEE Transactions on Signal Processing, pp.3445-62, 1993. (Transaction, journal).

[15] A. Moffat, R. Neal, I. H. Witten. "Arithmetic Coding Revisited", ACM Transactions on Information Systems, 16(3):256-294, 1998., (Journak paper)

**Dr.Hamid.A.Jalab:** Received his B.Sc degree from University of Technology, Baghdad, Iraq. MSc & Ph.D degrees from Odessa Polytechnic National State University 1987 and 1991, respectively. Presently, Visiting Senior Lecturer of Computer System and Technology, Faculty of Computer Science and Information Technology, University of Malaya, Malaysia. His areas of interest include neural networks and cryptography.

**Aos Alaa Zaidan**: He obtained his 1st Class Bachelor degree in Computer Engineering from university of Technology / Baghdad followed by master in data communication and computer network from University of Malaya. He led or member for many funded research projects and He has published more than 45 papers at various international and national conferences and journals, he has done many projects on Steganography for data hidden through different multimedia carriers image, video, audio, text, and non multimedia carrier unused area within exe. File, Cryptography and Stego-Analysis systems, currently he is working on the multi module for Steganography, Development & Implement a novel Skin Detector for increase the reliability. He is PhD Candidate on the Department of Electrical & Computer Engineering / Faculty of Engineering / Multimedia University / Cyberjaya, Malaysia. He is members IAENG, CSTA, WASET, and IACSIT. He is reviewer in the (IJSIS, IJCSNS, IJCSN, IJCSE and IJCIIS).

**Bilal Bahaa Zaidan:** He obtained his bachelor degree in Mathematics and Computer Application from Saddam University/Baghdad followed by master from Department of Computer System & Technology Department Faculty of Computer Science and Information Technology/University of Malaya /Kuala Lumpur/Malaysia, He led or member for many funded research projects and He has published more than 45 papers at various international and national conferences and journals. His research interest on Steganography & Cryptography with his group he has published many papers on data hidden through different multimedia carriers such as image, video, audio, text, and non multimedia careers such as unused area within exe. File, he has done projects on Stego-Analysis systems, currently he is working on multi module for Steganography, and he is PhD candidate on the Department of Electrical & Computer Engineering / Faculty of Engineering / Multimedia University / Cyberjaya, Malaysia, He is members IAENG, CSTA, WASET, and IACSIT. He is reviewer in the (IJSIS, IJCSNS, IJCSN, IJCSE and IJCIIS).